\documentclass[reprint,aps,prb,amsmath,amssymb,showpacs]{revtex4-1}
\usepackage{graphicx}% Include figure files
\usepackage{dcolumn}% Align table columns on decimal point
\usepackage{bm}% bold math
\usepackage{lineno}
\usepackage{hyperref}
\usepackage{epsfig,epstopdf,float,tabularx}
\usepackage[titletoc,title]{appendix}

\begin{document}
\renewcommand{\arraystretch}{1.8}

%\preprint{}

\title{Spin Coherence and $^{14}$N ESEEM Effects of Nitrogen-Vacancy Centers in Diamond with X-band Pulsed ESR}



\author{B. C. Rose$^1$}
%\email{brose@princeton.edu}
\author{C. D. Weis$^{2,3}$}
%\email{cdweis@lbl.gov}
\author{A. M. Tyryshkin$^1$}
%\email{atyryshk@princeton.edu}
\author{T. Schenkel$^2$}
%\email{t_schenkel@lbl.gov}
\author{S. A. Lyon$^1$}
%\email{lyon@princeton.edu}
\affiliation{$^1$Dept.\ of Electrical Engineering, Princeton University, Princeton, NJ 08544, USA}
\affiliation{$^2$Accelerator and Fusion Research Division, Lawrence Berkeley National Laboratory, Berkeley, California 94720, USA}
\affiliation{$^3$Dept. of Micro- and Nanoelectronic Systems, Ilmenau University of Technology, 98684 Ilmenau, Germany}


\date{\today}

\begin{abstract}
Pulsed ESR experiments are reported for ensembles of negatively-charged nitrogen-vacancy centers (NV$^-$) in diamonds at X-band magnetic fields (280-400 mT) and low temperatures (2-70 K).  The NV$^-$ centers in synthetic type IIb diamonds (nitrogen impurity concentration $<1$~ppm) are prepared with bulk concentrations of $2\cdot 10^{13}$ cm$^{-3}$ to $4\cdot 10^{14}$ cm$^{-3}$ by high-energy electron irradiation and subsequent annealing.  We find that a proper post-radiation anneal (1000$^\circ$C for 60 mins) is critically important to repair the radiation damage and to recover long electron spin coherence times for NV$^-$s.  After the annealing, spin coherence times of T$_2 = 0.74$~ms at 5~K are achieved, being only limited by $^{13}$C nuclear spectral diffusion in natural abundance diamonds. At X-band magnetic fields, strong electron spin echo envelope modulation (ESEEM) is observed originating from the central $^{14}$N nucleus. The ESEEM spectral analysis allows for accurate determination of the $^{14}$N nuclear hypefine and quadrupole tensors.  In addition, the ESEEM effects from two proximal $^{13}$C sites (second-nearest neighbor and fourth-nearest neighbor) are resolved and the respective $^{13}$C hyperfine coupling constants are extracted.
\end{abstract}


\maketitle

\section{\label{sec:intro}Introduction}
Nitrogen-vacancy centers (NV$^-$) in diamond are a type of point defect that have been widely studied for their spin dependent optical cycle\cite{Buckley2010}, long spin coherence at room temperature\cite{Gaebel2006}, and sensitivity to small fluctuations in magnetic\cite{Pham2011, Grinolds2011} and electric fields\cite{Dolde2011, Pham2011}.  They are presently a leading candidate for nanoscale magnetometry\cite{MazeMag2008, TaylorMag2008} and offer a promising platform for quantum computation\cite{2006WrachtrupQC}.  Coherent manipulation of single spins has been demonstrated by several groups\cite{Childress2006, Grinolds2011} and coherent information transfer with the central $^{14}$N nucleus and nearby $^{13}$C has been achieved\cite{Childress2006, Fuchs2011}.  Accurate control of the electron and nuclear spins requires the precise determination of the static spin Hamiltonian of the NV$^-$ and also understanding the environmental contributions to the spin decoherence rates.  In the first part of this manuscript we characterize electron spin coherence times at X-band magnetic fields.  We observe a strong effect of damage from electron irradiation on T$_2$ and show that appropriate annealing recovers a long T$_2=0.74$~ms limited by $^{13}$C spectral diffusion.  In the second part we report the electron spin echo envelope modulation (ESEEM) arising from the central $^{14}$N nucleus and deduce an accurate estimate of the $^{14}$N hyperfine and quadrupole tensors. The ESEEM effects from two proximal $^{13}$C sites are also resolved and the hyperfine couplings are derived.

Several mechanisms of NV$^-$ spin decoherence have been identified.  In high purity natural diamond the electron spin decoherence is set by spectral diffusion from the $1.1\%$ natural abundance of $^{13}$C nuclei and decoherence times T$_2=0.6$~ms have been observed for both single spins and ensembles of NV$^-$ at room temperature\cite{Stanwix2010,Mizuochi2009}.  The decoherence times were reported to be much shorter in heavily doped diamonds (nitrogen impurity, P1 center concentration $\sim$100 ppm) being limited by spectral diffusion from nitrogen impurities\cite{Takahashi2008}.  Damage from high energy processing techniques, like ion implantation, has been observed to drastically increase decoherence rates but can be repaired through proper high temperature annealing\cite{Naydenov2010,Yamamoto2013}.  Here we find that leftover damage from mild electron irradiation can also produce a strong effect on NV$^-$ decoherence rates and we suggest an annealing recipe which repairs this damage, restoring long T$_2 = 0.74$~ms, limited by the $^{13}$C nuclear flip-flops.  Previous T$_2$ measurements were made at low magnetic field ($<10$~mT) where strong ESEEM from distant $^{13}$C nuclei complicates the T$_2$ analysis\cite{Glasbeek1990,Childress2006,Stanwix2010}.  Here we use X-band magnetic fields where the distant $^{13}$C ESEEM is suppressed, allowing for accurate T$_2$ measurements and helping to resolve the radiation damage effects.

Weak ESEEM arising from the central $^{14}$N nucleus in NV$^-$ has recently been reported from ODMR experiments at low magnetic fields ($<20$~mT).\cite{Shin2014} We find that the $^{14}$N modulation is strongly enhanced at X-band magnetic fields (280-400~mT), allowing accurate determination of the nitrogen hyperfine and quadrupole tensors. 

\section{\label{sec:SetUp}Experimental Details}

Four diamond samples were used in our experiments (Table~\ref{table:SampTab}). All four are synthetic type IIb diamonds (ElementSix, CVD grown, P2 grade, [N] $<$ 1 ppm), with a concentration of NV$^-$ centers in the pristine material less than $10^{13}$ cm$^{-3}$ (below the detectable limit in our experiments). The samples were irradiated with electrons at 2 MeV energy, receiving a dose of $10^{15}$ cm$^{-2}$ (Sample A) and $10^{17}$ cm$^{-2}$ (Samples B, C, and D).  All samples were then annealed in a nitrogen atmosphere at $900^\circ$C for 20 minutes.  Samples C and D received an additional anneal in forming gas at $1000^\circ$C for 60 minutes.  Resulting concentrations of NV$^-$ centers in each sample were determined by comparing the ESR signal intensity to a standard sample with a known spin concentration (Table~\ref{table:SampTab}). Samples C and D received the identical radiation/annealing treatment and have the same concentration of NV$^-$ centers; the only difference between these two samples is their orientation (cut edges along $\left\{100\right\}$ in sample C versus $\left\{110\right\}$ in sample D) that allows different orientations of the crystals with respect to the external magnetic field in the ESR resonator. 

\begin{table*}[t]
\caption{Diamond samples used in this work with their electron irradiation and annealing steps. All samples are synthetic type IIb diamonds (ElementSix, CVD grown, P2 grade) with natural abundance (1.1\%) of $^{13}$C isotopes, total nitrogen impurity concentration of $< 1$~ppm, and paramagnetic neutral nitrogen (P1) center concentration of 0.1~ppm as estimated from ESR spin-counting.}
\begin{tabular}{c|c|c|c|c|c}
\hline\hline
Sample & Crystal edge & $e^{-}$ Fluence & \multicolumn{2}{c|}{Annealing recipe}  & [NV$^-$]\\
\cline{4-5}
label     &  orientation   & ($1/$cm$^2$)     & Temperature & Time              & ($1/$cm$^3$)\\[1pt]
\hline\hline
A & $\left\{100\right\}$ & $10^{15}$ & $900^\circ$C  & 20~min & $2\cdot10^{13}$\\
B & $\left\{100\right\}$ & $10^{17}$ & $900^\circ$C  & 20~min & $5\cdot10^{13}$\\
C & $\left\{100\right\}$ & $10^{17}$ & $1000^\circ$C & 60~min & $4\cdot10^{14}$\\
D & $\left\{110\right\}$ & $10^{17}$ & $1000^\circ$C & 60~min & $4\cdot10^{14}$\\
\hline\hline
\end{tabular}
\label{table:SampTab}
\end{table*}

X-band (9.6~GHz) pulsed ESR experiments were performed with a Bruker ESR spectrometer (Elexsys E580) using a dielectric resonator (ER-4118X-MD5) in a helium-flow cryostat (Oxford CF935). In most of the experiments a standard two-pulse (Hahn) echo pulse sequence ($\pi/2 - \tau - \pi - \tau - $echo) was used with $\pi/2$ and $\pi$ pulses set to be 50~ns and 100~ns.  

A frequency-doubled YLF laser (Spectra Physics, TFR-104Q-10), operating at 523~nm, was used for optical spin polarization of the NV$^-$ centers.  This Q-switched laser supplies 7~ns pulses with 200~$\mu$J per pulse at a 4~kHz repetition rate. The laser pulses were transmitted to the diamond samples inside the cold cryostat through a 4~mm diameter quartz rod used as an optical waveguide.  Typically between 10 to 100 laser pulses were sufficient to achieve steady-state spin polarization of the NV$^-$ centers in our experiments.

All ESR and ESEEM simulations discussed in this work were performed using the EasySpin toolbox developed for Matlab~\cite{EZSpin}. 

\section{\label{sec:ESR} ESR Spectra of NV$^-$ Centers at X-Band}

\begin{figure}[t]%NEED h and NOT H ...
\includegraphics[width=\linewidth]{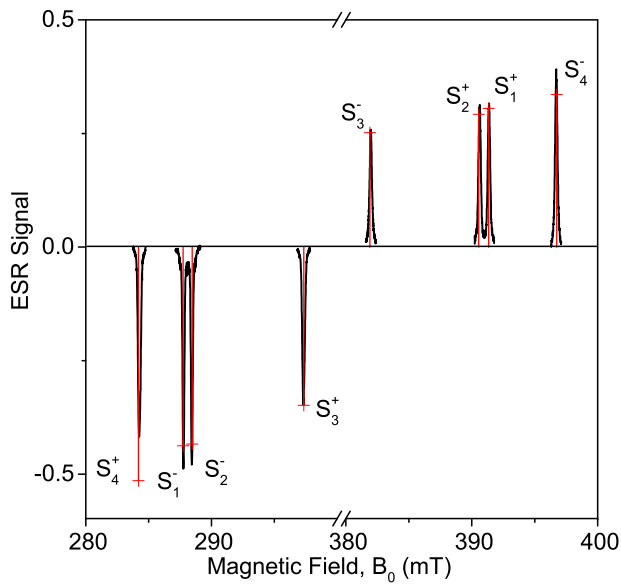}
\caption{(Color online) Experimental ESR spectrum (black peaks) of NV$^-$ centers in diamond (Sample~C) with magnetic field, B$_0$, oriented closely to the [110] crystal axis, measured at 10~K. The labels, S$^{\pm}_i$, assign the peak transitions to four non-equivalent NV$^-$ crystal sites ($i=1-4$) in diamond lattice and the $\pm$ sign identifies the transitions between $T_0\leftrightarrow T_+$ and $T_-\leftrightarrow T_0$, respectively.  Positive and negative amplitudes of the ESR peaks reflect non-equilibrium spin polarizations as resulted after an optical laser pumping. Red vertical sticks show the simulated positions of the ESR peaks using the Hamiltonian in Eq.~(\ref{eq:ESR}).  From this simulation, the magnetic field orientation B$_0$ was determined to be slightly misaligned from the intended [110] direction by ($2,2.2,0$)~degrees (Euler angles ZYZ).}
\label{fig:PolG} 
\end{figure}

ESR experiments were performed at three crystal orientations, with the external magnetic field (B$_0$) oriented closely along the [001], [110] and [111] axes of the diamond crystals.  Fig.~\ref{fig:PolG} shows an example of the ESR spectrum measured for Sample~C when the magnetic field (B$_0$) is oriented along [110].  Eight ESR peaks, labeled as S$^{\pm}_i$ in Fig.~\ref{fig:PolG}, correspond to four non-equivalent NV$^-$ crystal sites ($i=1-4$), and the $\pm$ sign identifies the transitions between $T_+\leftrightarrow T_0$ and $T_0 \leftrightarrow T_-$ for each site. Note that in this manuscript the spin eigenstates ($T_+$, $T_0$ and $T_-$) are defined in the {\it laboratory frame} with the spin quantization (Z) axis directed along the applied B$_0$ field.  This laboratory frame is natural in X-band experiments because the Zeeman interaction is the largest term in the NV$^-$ spin-Hamiltonian, in particular it is much larger than the zero field splitting (ZFS) term.  Our definitions of the spin eigenstates are thus different from those commonly used in low magnetic field experiments where the Z-axis is associated with the orientation of the ZFS principal axis (the {\it molecular frame}).

In each experiment the crystal orientation was determined with sub-degree accuracy by fitting the measured positions of all eight  NV$^-$ ESR peaks using the spin Hamiltonian: 
\begin{equation}
\hat{H_0}=\mu_B \text{B}_0\hat{\textbf{g}}{\textbf{S}}+{\textbf{S}}\hat{\textbf{D}}{\textbf{S}},
\label{eq:ESR}
\end{equation}
where $\mu_B$ is a Bohr magneton, and \textbf{S} is an electron spin (S$=1$) vector operator. In simulations we assumed the ZFS tensor ($\hat{\textbf{D}}$) to be axial with $D=2.873$ GHz and the electron g-tensor ($\hat{\textbf{g}}$) to be isotropic with $g = 2.0030$ as expected for the $C_{3v}$ symmetry of NV$^-$.\cite{Felton2009}  Thus, from the ESR peak position simulations in Fig. ~\ref{fig:PolG} we determined that the magnetic field vector (\textbf{B}$_0$) was slightly misaligned from the intended [110] by $(\alpha, \beta,\gamma)= (2,2.2,0)$~degrees where the three angles are Euler rotations (ZYZ) from [110]. 

Positive (emission) and negative (absorption) amplitudes of the ESR peaks in Fig.~\ref{fig:PolG} reflect non-thermal triplet state populations after optical pumping.  Four non-equivalent NV$^-$ crystal sites can be subdivided into two groups. Two sites (1 and 2) show positive S$^{+}_i$ peaks and negative S$^{-}_i$ peaks indicating a preferential $T_0$ state population during the optical pumping.  Two other sites (3 and 4) show the opposite signs of the S$^{\pm}_i$ peaks revealing a preferential $T_{\pm}$ state's population.  This preferential $T_{\pm}$ state population is in stark contrast to the previously reported preferential $T_0$ state population as observed in low magnetic field experiments (B$_0 < 50$~mT).\cite{Robledo2011} We find that at X-band magnetic fields (B$_0 = 200-400$~mT) the optically-induced triplet state population strongly depends on the orientation of B$_0$ with respect to the principal ZFS axis of NV$^-$ (associated with the N-V bond direction).  Two NV$^-$ sites (1 and 2) in Fig.~\ref{fig:PolG} have their principal ZFS axes oriented at a 90 degree angle with respect to B$_0$ and therefore preferential $T_{0}$ state population during optical pumping. The two other sites (3 and 4) have their angle at approximately 35 degrees and preferential $T_{\pm}$ state population.  Further details of the B$_0$ orientation and magnitude dependence for laser optical polarization in NV$^-$ centers will be published separately.\cite{Brendon2016} 

\section{\label{sec:T2} Electron Spin Coherence of NV$^-$}

\begin{figure}[t]%NEED h and NOT H ...
\includegraphics[width=\linewidth]{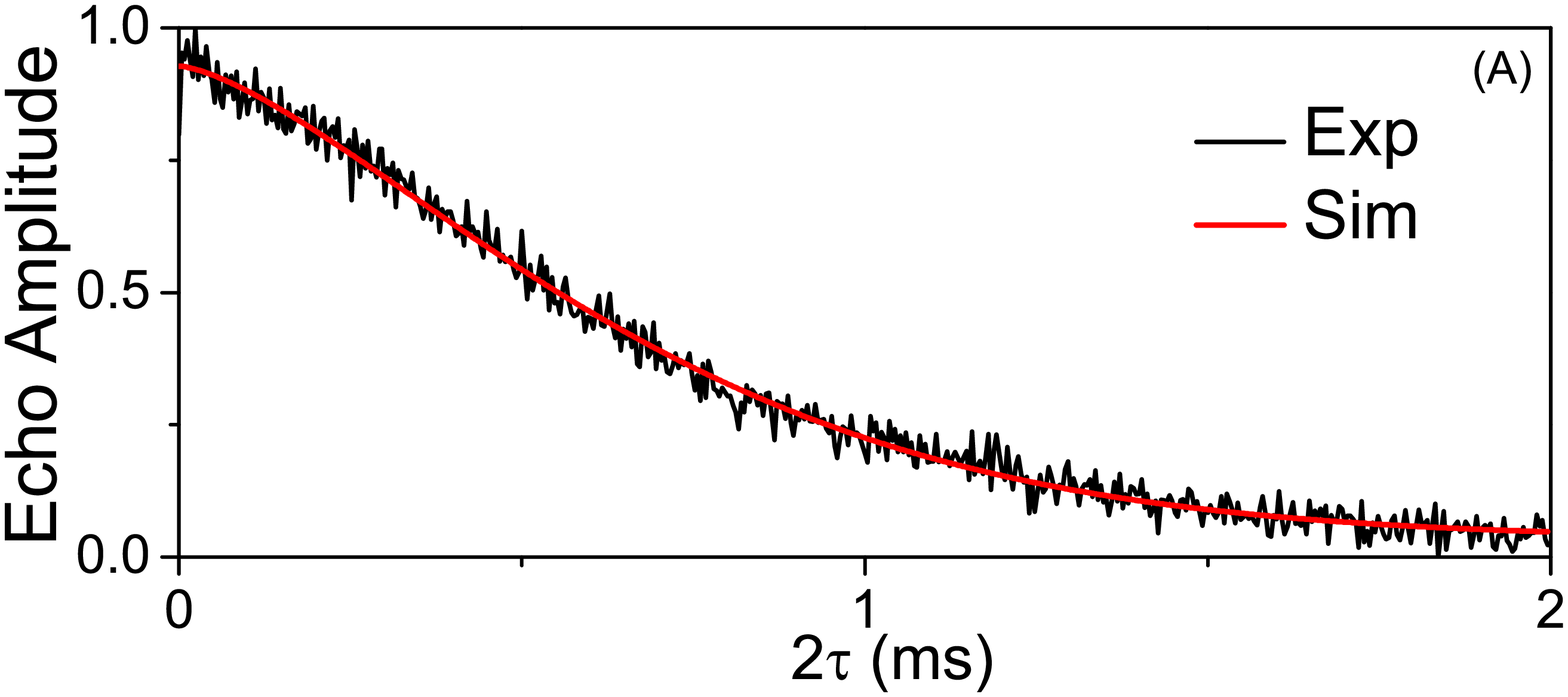}
\includegraphics[width=\linewidth]{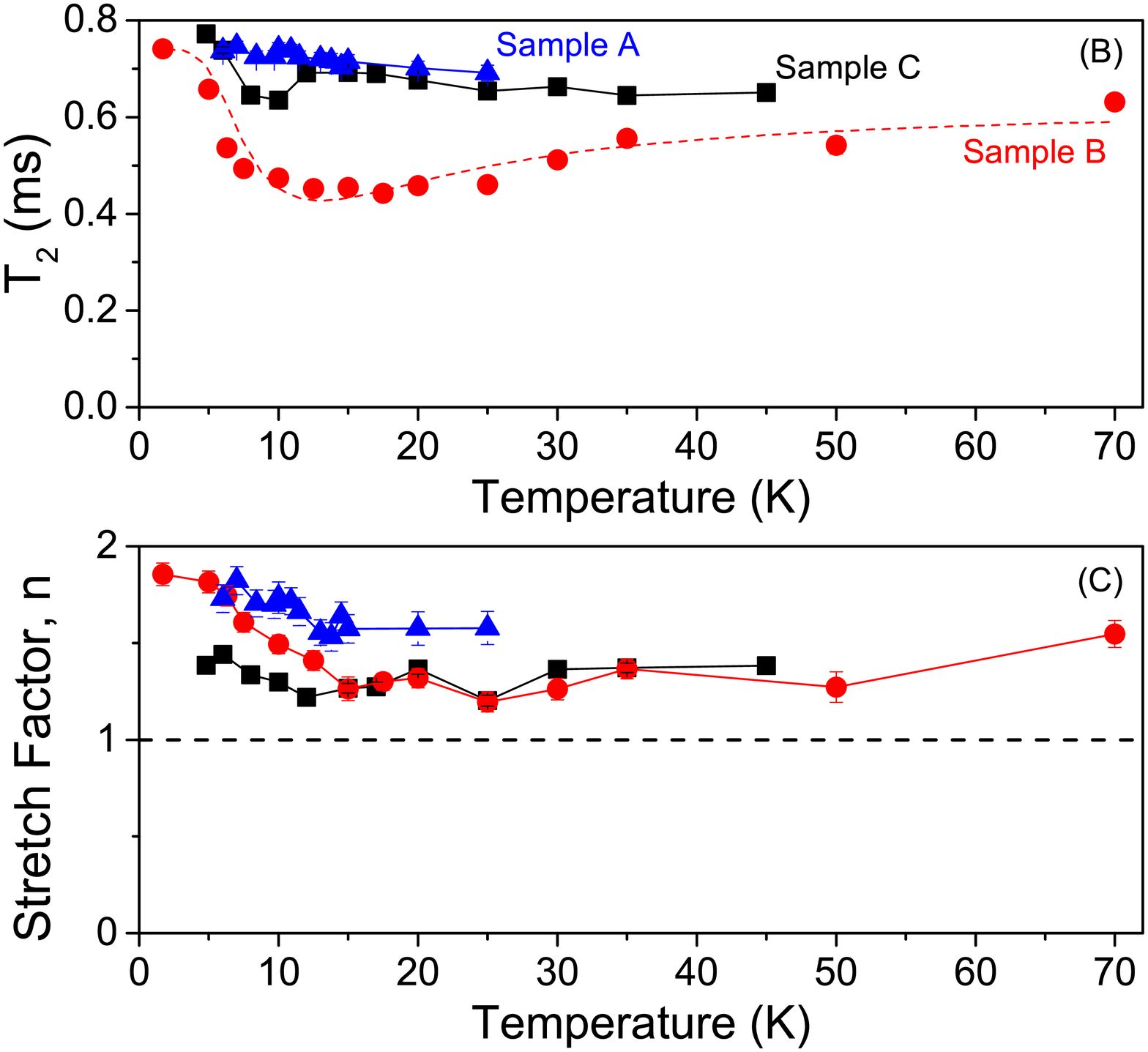}
\caption{(Color online) (A) A two-pulse (Hahn) echo decay of NV$^-$ centers in Sample~C measured at 6~K and B$_0 \parallel [110]$. The ESR peak at B$_0=390$~mT, marked as S$^+_2$ in Figure \ref{fig:PolG}, was used in this experiment.
The red solid curve is a fit using Eq.~(\ref{eq:T2decay}) with T$_2=0.74\pm 0.01$~ms and $n=1.45\pm0.02$.  Non-exponential decay is due to spectral diffusion from $^{13}$C nuclear spin flip-flops.  (B and C) Temperature dependences of coherence time (T$_2$) and exponential factor ($n$) for Samples A--C measured on the S$^+_2$ peak with B$_0\parallel [110]$. Larger $\bar{e}$ fluence in Sample~B results in significant reduction of T$_2$ as compared to Sample~A. Subsequent annealing in forming gas at $1000^\circ$C for 60 minutes (Sample~C) repairs the radiation damage and restores T$_2$ back to the level seen in Sample A.  The dashed red curve in (B) is a fit to Sample B's temperature dependence assuming a bath of thermally activated magnetic field fluctuators with a single activation energy of 2.5 meV.}
\label{fig:AnnealG}
\end{figure}

Two-pulse Hahn echo experiments were performed to measure NV$^-$ electron spin coherence times in Samples A--D covering the temperature range 2-70~K.  Figure~\ref{fig:AnnealG} presents the results for three samples A--C measured at magnetic field orientation $\textbf{B}_0\parallel [110]$ (Sample D is not shown but it gave results similar to Sample C).  Selected experiments were also performed at $\textbf{B}_0\parallel [001]$ and $\textbf{B}_0\parallel [111]$ resulting in the same coherence times within experimental errors. In all samples, and over the full temperature range, the decays were non-exponential as illustrated for Sample~C in Fig.~\ref{fig:AnnealG}(A). The decays can be fit using 
\begin{equation}
V(2\tau) = A \cdot \exp \left({-\left({2\tau}/{\text{T}_2}\right)^n}\right),
\label{eq:T2decay}
\end{equation}
where $\tau$ is the interpulse delay in the Hahn echo sequence.  Temperature dependences of the coherence time (T$_2$) and the stretch factor ($n$) were extracted from the decay fits for Samples~A--C and are summarized in Figure~\ref{fig:AnnealG}(B,C).  

\subsection{\label{sec:Tsd} NV$^-$ Decoherence due to $^{13}$C Nuclear Spin Flip-Flops}

We start our discussion with Samples A and C where we find T$_2$ and $n$ to be independent of temperature (less than 20\% variation) throughout the measured range 5 -- 25~K and 5 -- 45~K, respectively.  The measured T$_2$ = 0.7~ms are comparable in both samples, however the exponent, $n$, is substantially smaller in Sample C.  Three temperature-independent mechanisms are potentially responsible for NV$^-$ decoherence in these samples: (1) instantaneous diffusion due to $\pi$-pulse induced electron spin flips of nearby NV$^-$ centers,~\cite{Klauder1962,Milov1998} (2) spectral diffusion due to flip-flopping dipolar fields from other paramagnetic defects (mostly substitutional nitrogen P1 centers),\cite{Takahashi2008,Wang2013} and (3) spectral diffusion due to $^{13}$C nuclear spin flip-flops.~\cite{Maze2008,Stanwix2010}  The first mechanism (instantaneous diffusion) can  immediately be excluded after recognizing that T$_2$ in Sample C is comparable to that in Sample A even though the NV$^-$ concentration in Sample C is 20 times larger than in Sample A. An instantaneous diffusion decoherence rate ($1/$T$_2$) should scale proportionally with NV$^-$ concentration,\cite{Milov1998} and therefore T$_2$ should be 20 times as short in Sample C as compared to Sample A if it was limited by instantaneous diffusion.

The second mechanism (spectral diffusion from P1 center flip-flops) has been observed to be a dominant source of NV$^-$ decoherence in highly-doped diamond samples (20--200~ppm of nitrogen impurities), often limiting T$_2$ to a microsecond timescale.\cite{Takahashi2008,deLange2010} However, this mechanism is less important in our samples because of their smaller P1 concentrations (0.1~ppm).  A rough estimate of the expected contribution from this mechanism can be made using simulated results in diamond with high P1 concentrations (1--100~ppm)\cite{Takahashi2008} and extrapolating to our 0.1~ppm concentration.  In addition, we have to take into account the inhomogeneous broadening of P1 ESR transitions ($\sim 300$~kHz) due to $^{13}$C hyperfine interactions in natural diamonds.\cite{vanWyk1997b} The inhomogeneous broadening detunes the P1's from one another in a highly localized uncorrelated manner limiting the number of resonant flip-flopping pairs.\cite{Witzel2010}  For our [P1]~$= 0.1$~ppm, we estimate the average dipole-dipole coupling between P1 spins in pairs to be $\sim 1$~kHz which is 300 times smaller than the spin detuning from the inhomogeneous broadening so that only 1/300th of the total number of pairs are allowed to flip-flop.  The direct extrapolation from the results in \citet{Takahashi2008} to our P1 concentration 0.1~ppm gives T$_2 = 200\;\mu$s, and after accounting for the inhomogeneous broadening we estimate the contribution from P1 spin flip-flops to NV$^-$ decoherence in our sample to be only T$_2 = 60$~ms. This T$_2$ is much longer than the 0.74~ms measured in our experiments.
%The inhomogeneity effect has not been taken into account in the simulations of highly-doped diamonds\cite{Takahashi2008}, however it becomes important at our low [P1] densities. 

The effect of $^{13}$C nuclear spin flip-flops on NV$^-$ decoherence in natural diamonds (1.1\%  of $^{13}$C isotopes with $I=1/2$) has already been studied using low-field, room-temperature ODMR experiments for single spins as well as ensembles of spins, and T$_2=0.63$~ms was reported\cite{Stanwix2010, Maze2008}.  Their T$_2$ is slightly shorter but is similar to the 0.74~ms we measure in Samples A and C at high magnetic fields 280-400~mT and low temperatures 5-25~K.  We speculate that in both cases T$_2$ is dominated by spectral diffusion from $^{13}$C nuclear spin flip-flops and that this decoherence process is both temperature and field independent, as expected.  The measured n=1.75 in Sample~A is close to 2 as expected for nuclear-induced spectral diffusion.\cite{deSousa2003}  Note that for Sample~C, n$=1.4$ is noticeably smaller than 2, indicating a contribution from some other (unidentified) mechanism, possibly related to the mechanism discussed in the next section for Sample B.  T$_2$ is slightly shorter in the low-field ODMR experiments which could be due to residual interference from $^{13}$C modulation even at the [111] field orientation\cite{HahnESEEM, Maze2008, Stanwix2010}. We note that in our experiments, at higher magnetic fields (280-400~mT), the $^{13}$C ESEEM effects from distant nuclei are fully suppressed at all field orientations and we observe no orientation dependence in T$_2$ in contrast to the low field ODMR experiments.  

\subsection{\label{sec:Anneal} Effect of Electron Irradiation Damage and Post-Radiation Annealing on NV$^-$ Decoherence}

Sample B in Fig.~\ref{fig:AnnealG}(B) shows a distinctly different temperature dependence of  T$_2$ from Samples A and C. At high and low temperatures, T$_2$ approaches 0.7~ms, close to that in Samples A and C.  However, T$_2$ drops down at intermediate temperatures reaching a minimum of 0.45~ms (a 35\% reduction) at around 15~K.  This type of T$_2$ temperature dependence has been observed in situations where thermally activated electric or magnetic field noise dominates spin relaxation.~\cite{Mims1968,Hilczer1995,Eatons2000} As temperature decreases the characteristic time of this noise ($\tau_\text{c}$) increases, making a transition from a motional narrowing regime ($\tau_\text{c} \ll \text{T}_2$) at high temperatures to a slow spectral diffusion regime ($\tau_\text{c} \gg \text{T}_2$) at low temperatures. The effect of noise is minimal in these two extreme regimes (therefore long T$_2$), however the effect on T$_2$ can become significant in the intermediate regime when $\tau_\text{c} \sim \text{T}_2$.

The change in $\tau_\text{c}$ occurs over a relatively narrow temperature range (5--25 K) suggesting a thermally activated process.  While the nature of defect sites remain unknown, if we assume a process inhibited by a single energy barrier, e.g. $1/\tau_\text{c} \sim \exp(-E_a/kT)$, the temperature dependence of T$_2$ for Sample~B gives an estimate of the activation energy to be $E_a=2.5$~meV and a density of magnetic fluctuators to be 2.7$\cdot$10$^{16}$ cm$^{-3}$ as shown in Fig. \ref{fig:AnnealG}(B) (dashed red line).  Alternatively, the source of this decohering noise can be electric in origin.  Electric field noise from charge fluctuators modulates the ZFS of the NV$^-$ through a Stark shift, contributing to dephasing.  

Sample~B received a $100\times$ higher electron dose than Sample~A, resulting in a much higher density of damage-induced defects.  The 20~minute anneal at $900^\circ$C, was sufficient for repairing most of the damage defects in Sample~A, but was not sufficient for Sample~B.  A longer anneal (60~minutes) at a higher temperature ($1000^\circ$C) was required  in order to repair the excess radiation damage.  This is confirmed by the observed recovery of T$_2$ in Sample~C which was otherwise prepared the same as Sample~B.  

Similar reduction in T$_2$ was observed for NV$^-$ after high-energy ion implantation, with longer T$_2$ recovered after subsequent high temperature anneals.\cite{Naydenov2010,Yamamoto2013} It was argued that paramagnetic vacancy clusters are primarily responsible for short T$_2$ of NV$^-$s in the pre-annealed diamonds and that concentration of these vacancy clusters can be greatly reduced by annealing at $1000^\circ$C.

Despite the $100\times$ higher radiation dose, Sample~B shows only a $2.5\times$ increase in the number of NV$^-$ centers compared to Sample~A.  Annealing the sample longer at higher temperature further mobilizes vacancies allowing them to be trapped by substitutional nitrogen resulting in a higher concentration of NV$^-$ centers in Sample~C. The final annealing of Sample~C resulted in an  $8\times$ increase in NV$^-$ concentration compared to Sample~B, and a total of $20\times$ increase in density compared to sample~A.

To conclude, when fabricating NV$^-$ rich diamond via electron irradiation, the subsequent annealing step is of critical importance in order to achieve a high density of NV$^-$ centers and also to avoid additional spin decoherence.  

\section{\label{sec:ESEEM} $^{14}$N Nuclear Modulation Effects in NV$^-$}

\begin{figure*}[!t]%NEED h and NOT H ...
\includegraphics[width=\linewidth]{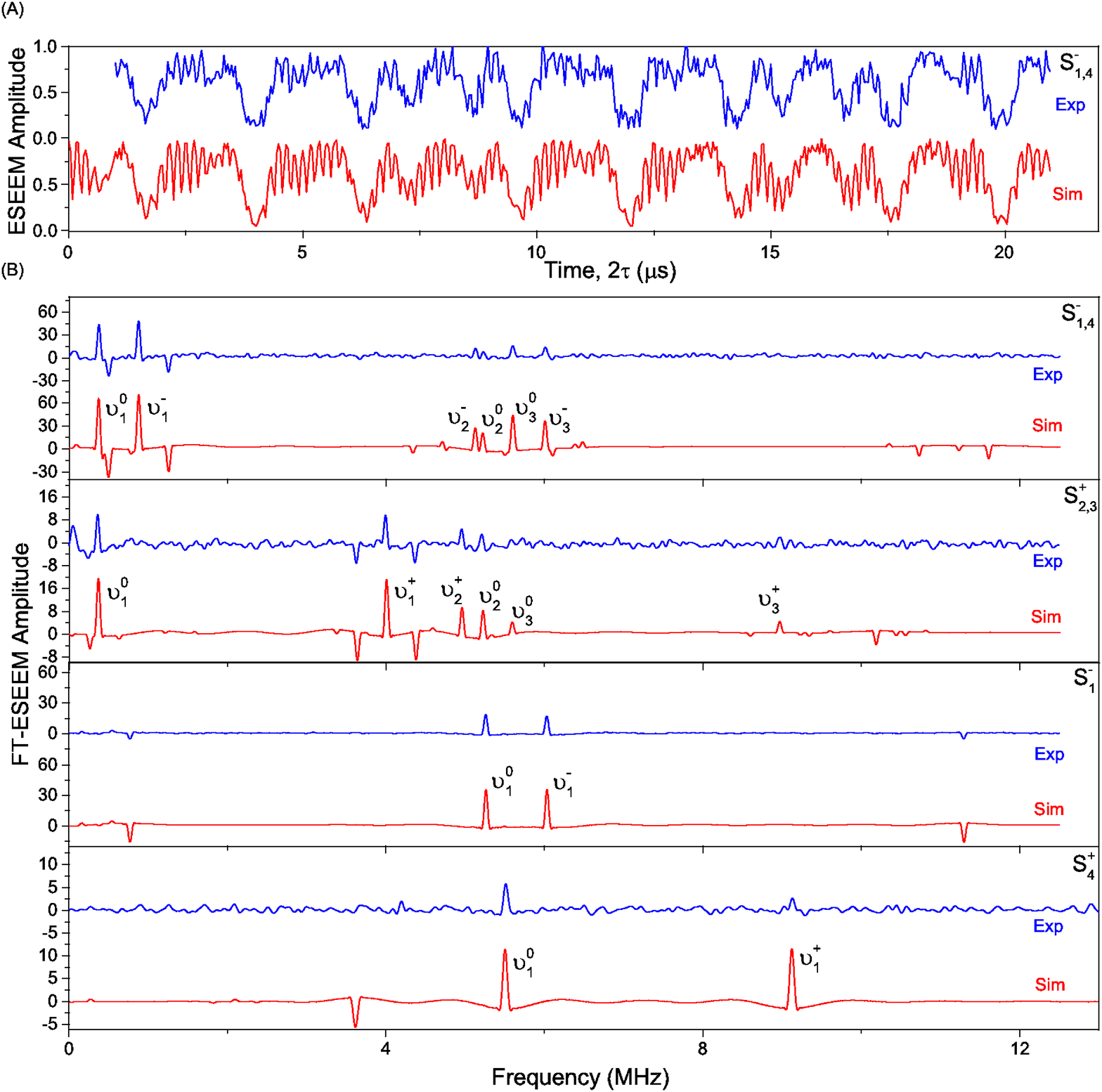}
\caption{(Color online) $^{14}$N ESEEM effects in a two-pulse (Hahn) echo experiment for NV$^-$ centers in Sample~C at 4.8~K.  (A) Experimental (blue) and simulated (red) ESEEM time-domains measured on the ESR peak S$^-_{1,4}$ with B$_0\parallel [001]$.  The experimental trace was normalized by the relaxation decay. Only the first 20~$\mu$s of the time-domain are shown, although the modulation extends for hundreds of microseconds with little modulation damping. Simulations were done with EasySpin~\cite{EZSpin} using the spin Hamiltonian of Eq.~\ref{eq:ESEEMH}.  (B) Experimental (blue) and simulated (red) cosine FT-ESEEM spectra for
S$^-_{1,4}$ and S$^+_{2,3}$ at B$_0\parallel [001]$, and for S$^-_{1}$ and S$^+_{4}$ at B$_0\parallel [110]$. The peaks in the FT spectra are labeled with $\nu_{i}^{0,\pm}$ identifying three $^{14}$N nuclear spin transitions for each electron spin state $T_{0,\pm}$.
Experimental peak intensities are smaller than the simulation at high frequencies due to the finite detection bandwidth used in our experiments.}
\label{fig:n0B-ESEEM}
\end{figure*}

While measuring two-pulse (Hahn) echo decays we observed strong nuclear modulation effects (ESEEM),\cite{Dikanov1992} superimposed on the decays.  The modulated echo signals and their respective Fourier-Transform (FT) spectra are shown in Figs.~\ref{fig:n0B-ESEEM}-\ref{fig:npar111-ESEEM}. The modulation effects are most pronounced when measuring the S$^-_\text{1-4}$ transitions (the $T_-\leftrightarrow T_0$ transitions) with the magnetic field oriented along [001] (Fig.~\ref{fig:n0B-ESEEM}A).  On the other hand, the modulation effects are strongly suppressed for the S$^+_{3,4}$ transitions ($T_0\leftrightarrow T_+$) with the magnetic field oriented along [110] (it is more clearly seen in the time-domain traces shown in Fig.~\ref{fig:TimeDomains} in Appendix).  When B$_0\parallel[111]$, the $^{14}$N modulation is completely suppressed on both S$^{\pm}_{1}$ transitions for the NV$^-$ centers whose symmetry axis is directed along the magnetic field (Fig.~\ref{fig:npar111-ESEEM}). At this field orientation, a weak $^{13}$C modulation is observed resulting from proximal $^{13}$C nuclear spins as further discussed in Sec.~\ref{sec:ESEEM_13C}. The transitions with minimum modulation effects were used in our T$_2$ measurements in Sec.~\ref{sec:Tsd} in order to minimize the distortion of the measured T$_2$ times.

The Fourier transform spectra (FT-ESEEM) in Fig.~\ref{fig:n0B-ESEEM}B are straightforward to interpret. The peaks with positive amplitudes, known as 'basic' ESEEM harmonics, arise from $^{14}$N nuclear spin transitions within the two corresponding electron spin manifolds being driven. For example,  the FT-ESEEM spectrum measured while driving the $T_-\leftrightarrow T_0$ transition in Fig.~\ref{fig:n0B-ESEEM}B(top) reveals the nuclear spin transitions within the $T_-$ and $T_0$ electronic spin manifolds. The ESEEM peaks here are labeled $\nu_i^-$ and $\nu_i^0$ ($i=1..3$).  Similarily, in the second from top spectrum in Fig.~\ref{fig:n0B-ESEEM}B, the nuclear spin transitions are within the $T_0$ and $T_+$ electron spin manifolds (labeled $\nu_i^0$ and $\nu_i^+$). Within each group the three nuclear spin transitions satisfy the additive relationship, $\nu_1^j + \nu_2^j = \nu_3^j$, as expected for the three transitions between nuclear spin states of I~$=1$. Finally, the ESEEM peaks with negative amplitudes, known as 'combination' ESEEM harmonics, are sum and difference combinations of the basic harmonics. 

The FT-ESEEM spectra measured at B$_0 \parallel [110]$ (two bottom spectra in Fig.~\ref{fig:n0B-ESEEM}B) show fewer basic harmonics (only one harmonic for each $T_{0,\pm}$ state involved).  Some electron-nuclear flip-flop transitions (known as forbidden or branching transitions)\cite{Dikanov1992,SchweigerBook} are forbidden at this field orientation which explains the smaller number of harmonics and also the shallower $^{14}$N modulation.  When B$_0 \parallel [111]$ (Fig.~\ref{fig:npar111-ESEEM}), the branching transitions are completely forbidden for NV$^-$ centers whose symmetry axis is parallel to B$_0$.  These specific NV$^-$ centers have their (axial) hyperfine and quadrupolar tensors directed along B$_{0}$, and therefore the $C_{3v}$ symmetry of the NV$^-$ site is preserved even with the field applied.  In this situation the $^{14}$N nuclear eigenstates are the same for all three $T_{0,\pm}$ electron spin states, and therefore nuclear spin state branching when applying a $\pi$ pulse in a Hahn echo is forbidden and no $^{14}$N modulation effects are observed.\cite{Dikanov1992,SchweigerBook}

We simulate the $^{14}$N ESEEM effects using the following spin Hamiltonian:
\begin{equation}
\hat{H_0}=\mu_B \textbf{B}_0\hat{\textbf{g}}{\textbf{S}}+{\textbf{S}}\hat{\textbf{D}}{\textbf{S}}+
\textbf{S}\hat{\textbf{A}}\textbf{I}+\textbf{I}\hat{\textbf{Q}}\textbf{I}-g_{n}\mu_n\textbf{B}_0\textbf{I}.
\label{eq:ESEEMH}
\end{equation}
In addition to the electron Zeeman and ZFS terms needed to describe ESR transitions, we include here the nuclear hyperfine coupling ($\hat{\textbf{A}}$), nuclear quadrupole coupling ($\hat{\textbf{Q}}$), and the nuclear Zeeman terms ($g_n$ and $\mu_n$ are the nuclear $g$-factor and the nuclear magneton, respectively).  The hyperfine and nuclear quadrupolar tensors are assumed to be axial, and both tensors and also the ZFS tensor are assumed to be coaxial with the principal axes directed along the N$-$V bond. The latter is required by the $C_{3v}$ symmetry of the NV$^-$ defect.  

In each ESEEM experiment the orientation of the magnetic field (B$_0$) with respect to the crystal axes was determined by simulations of the observed ESR peak positions (similar to the fits shown in Fig.~\ref{fig:PolG}).  In terms of three Euler angle rotations ($\alpha,\beta,\gamma$), the magnetic field orientations were misaligned by ($8,1,0$)~degrees from [001] in the two top experiments shown in Fig.~\ref{fig:n0B-ESEEM}, by ($1.1,2.1,0$)~degrees from [110] in the two bottom experiments shown in Fig.~\ref{fig:n0B-ESEEM}, and by ($0.3,0.9,0$)~degrees from [111] in the experiments shown in Fig.~\ref{fig:npar111-ESEEM}.  Before Fourier transformation the initial portion of the simulated time-domain traces ($\tau < 0.5~\mu$s) was removed in order to account for the experimental dead time.   The phase skew in the FT spectra introduced by the dead time was adjusted using first order phase correction.

\begin{table}[t]
\caption{$^{14}$N hyperfine and nuclear quadrupolar coupling parameters (in MHz) derived from our ESEEM data and their comparison with previously reported ESR and ENDOR results.}
\begin{tabular}{lccc}
\hline\hline
Source\hspace{2cm} & A$_\parallel$ & A$_\perp$ & P$_\parallel$\footnote{P$_\parallel = \frac{3}{4} \frac{e^2Qq}{\hbar}$, where $Q$ is an electric quadrupole moment of $^{14}$N nucleus, and $q$ is the electric field gradient at the nucleus. The rhombicity parameter $\eta$ was assumed to be zero for an axially symmetric nuclear quadrupole tensor of NV$^-$ center.}\\
\hline
This work                              &-2.19(2)&-2.65(3)&-4.95(2)\\
Felton et al.\cite{Felton2009}  &-2.14(7)&-2.70(7)&-5.01(6)\\
He et al.\cite{He1993}           &+2.30(2)&+2.10(10)&-5.04(5)\\
\hline
\hline
\label{table:NucRes}
\end{tabular}
\end{table}

The numerical simulations (shown in red in Figs.~\ref{fig:n0B-ESEEM}) are in excellent agreement with experimental data for B$_0\parallel [100]$ and [110].  Both the peak positions and amplitudes in the FT spectra as well as the modulation features observed in the time-domain traces (Fig.~\ref{fig:TimeDomains} in Appendix) are clearly reproduced.  The simulations produce slightly larger modulation amplitudes than experiment, especially for the high-frequency ESEEM peaks, because they do not take into account the limited detection bandwidth used in our experiments (the bandwidth was set by a 100~ns integration window). The simulation also show no $^{14}$N modulation at B$_0\parallel$ [111] for the NV$^-$ sites with their axis parallel to B$_0$ as observed in the experiment (Figs.~\ref{fig:npar111-ESEEM}).

The extracted $^{14}$N hyperfine and nuclear quadrupole parameters are summarized in Table~\ref{table:NucRes}, where they are also compared with the previous results from the ESR/ENDOR experiments. Our parameters agree closely with those reported by Felton et al.,\cite{Felton2009} however we significantly improve on the accuracy of the parameters due to the high resolution of our FT-ESEEM spectra as compared to their ESR/ENDOR spectra. Our parameter errors reported in Table~\ref{table:NucRes} are dominated by errors in orienting the crystals during the experiments.

Nuclear modulation effects are expected to be greatly enhanced when the {\it cancellation} condition is met.\cite{Dikanov1992,SchweigerBook} In the case of NV$^-$ ($\text{S}=1$ and $\text{I}=1$), this condition occurs when $|\nu_I-m_S\text{A}|<|\text{Q}|$, which is when the Zeeman energy ($\nu_I=g_n\mu_n\text{B}_0$) is approximately canceled by the hyperfine energy (A) for one of the electron spin projections $m_S$. For a $^{14}$N hyperfine coupling of $\text{A}\sim-2$~MHz for NV$^-$ (Table~\ref{table:NucRes}) this condition is met when B$_0\sim350$~mT for the $T_-$ state ($m_S=-1$).  This field is close to the fields used in our [001] experiments which explains the deep modulation (modulation depth $\sim 95\%$) observed for the $T_- \leftrightarrow T_0$ transitions (S$^-_{1,4}$ in Fig.~\ref{fig:n0B-ESEEM}, top spectrum). Somewhat weaker modulation (modulation depth $\sim 20-50\%$) observed for the $T_+ \leftrightarrow T_0$ transitions (S$^+_{2,3}$ in Fig.~\ref{fig:n0B-ESEEM}, second top spectrum) and also for both $T_{\pm} \leftrightarrow T_0$ transitions at [110] (S$^\pm_i$ in Fig.~\ref{fig:n0B-ESEEM}, two bottom spectra) is explained by B$_0$ being further away from the cancellation condition.  Weak modulation from the central $^{14}$N nucleus in NV$^-$ has been recently reported in low-field ODMR experiments\cite{Shin2014}. The $^{14}$N modulation was weak with modulation depth $\sim 3\%$ in those experiments because the magnetic field (B$_0=7.5$~mT) was far away from the cancellation condition.  The Fourier spectral components were difficult to resolve accurately under these conditions.

\section{\label{sec:ESEEM_13C}  Nuclear Modulation Effects from Proximal $^{13}$C Nuclei in NV$^-$}
The ESEEM effects resulting from hyperfine interactions with $^{13}$C nuclear spins have been previously reported for NV$^-$ centers in natural diamonds in ODMR experiments at low magnetic fields (B$_0<20$~mT).~\cite{Glasbeek1990,VanOort1990,Maze2008,Stanwix2010} At these fields the modulation effects were shown to be due to {\it distant} $^{13}$C nuclear spins ranging from 0.5-1.6~nm (4-10 lattice sites away) from the NV$^-$ center.\cite{VanOort1990} 
These distant $^{13}$C modulation effects are completely suppressed in our X-band experiments (B$_0 = 280-400$~mT) because the weak hyperfine couplings to distant $^{13}$C spins ($\text{A}<0.1$~MHz) are now far from the cancellation condition. We note that in the case of an electron spin $\text{S}=1$ and a $^{13}$C nuclear spin $\text{I}=1/2$, the cancellation condition occurs when $|\nu_I-m_S\text{A}|< \text{T}$, where T is an anisotropic part of the $^{13}$C hyperfine coupling.\cite{Dikanov1992,SchweigerBook}

\begin{figure}[!t]
\includegraphics[width=\linewidth]{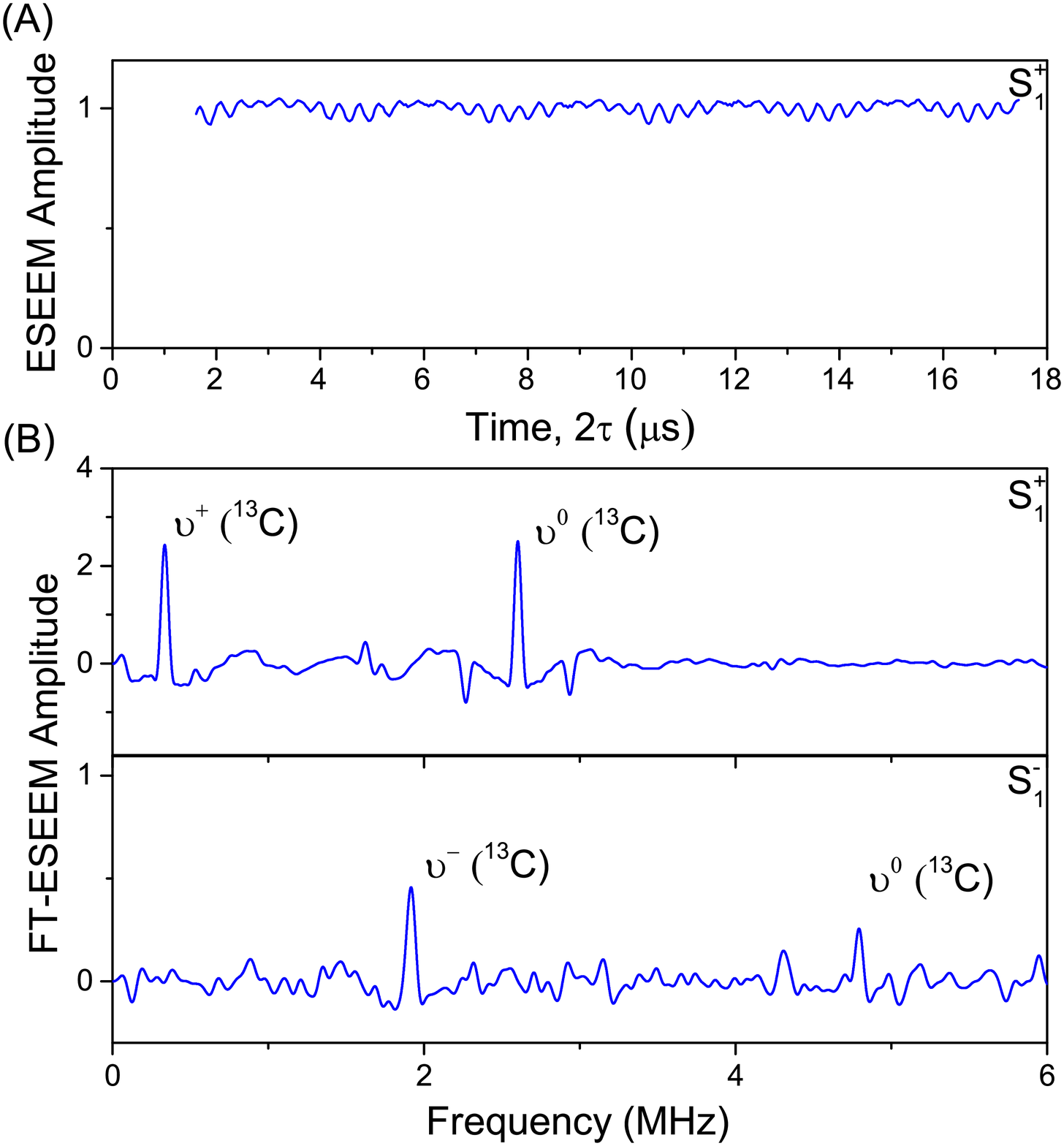}
\caption{(Color online)  Experimental $^{13}$C ESEEM time-domains (A) and FT spectra (B) measured using 2-pulse (Hahn) echo experiment in Sample~D at 4.8~K and B$_0\parallel [111]$.  The traces shown are for the S$^+_1$ and S$^-_1$ transitions of NV$^-$ defects whose symmetry axis (the N$-$V bond) is oriented parallel to the applied magnetic field B$_0$. $^{14}$N modulation effects are suppressed at this field orientation because of the $C_{3v}$ symmetry of NV$^-$ defects. Instead, a weak $^{13}$C modulation is resolved arising from carbon nuclear spins located at the second and fourth nearest-neighbor lattice sites around NV$^-$'s. The $^{13}$C peaks are labeled with $\nu^{0,\pm}$ in accordance with their electronic spin state, $T_{0,\pm}$.}
\label{fig:npar111-ESEEM}
\end{figure}

In our strong magnetic fields (B$_0 = 280-400$~mT) the cancellation condition is met for the {\it proximal} $^{13}$C spins as evidenced by the modulation effects in Fig.~\ref{fig:npar111-ESEEM}.  The peaks in the FT spectra Fig.~\ref{fig:npar111-ESEEM}B are labeled with $\nu^{0,\pm}$ identifying the related electron spin states $m_S={0,\pm1}$.  In each spectrum the peaks $\nu^{0}$ appear at exactly the $^{13}$C nuclear Zeeman frequency ($\nu_I=g_n\mu_n\text{B}_0$), and the peaks  $\nu^{\pm}=|\nu_I\mp\text{A}|$ are shifted away by the hyperfine coupling.  Thus, the hyperfine coupling constants for the proximal $^{13}$C spins can immediately be estimated from the observed peak positions. From the top spectrum in Fig.~\ref{fig:npar111-ESEEM} we estimate $\text{A}_1=2.56(2)$ or 2.92(2)~MHz, where two  values are possible because of the uncertainty in the sign of $(\nu_I\mp\text{A}_1)$.  From the bottom spectrum, we estimate $\text{A}_2=-2.88(4)$ or $-6.70(4)$~MHz.  The two estimated hyperfine coupling constants ($\text{A}_1$ and $\text{A}_2$) are different in both magnitude and sign, indicating that they must belong to two distinct $^{13}$C lattice sites in close proximity to the NV$^-$ center.

Hyperfine coupling constants for several $^{13}$C lattice sites around the NV$^-$ have recently been reported in low-field ODMR experiments and assigned to specific sites using density functional theory calculations.\cite{Gali2008,Smeltzer2011,Dreau2012}  Our $\text{A}_1=2.56$~MHz agrees well with their 2.54~MHz measured for one of the fourth nearest-neighbor lattice sites ("Site G" according to the notation introduced in Ref.~[\cite{Smeltzer2011}]), while our $\text{A}_2=-6.7$~MHz is close to their -6.55~MHz measured for a second nearest-neighbor site ("Site D").  Each of these sites ("G" and "D") involves six equivalent (symmetry-related) lattice positions around the NV$^-$. Therefore, even though the isotopic abundance of $^{13}$C is only 1.07\%, the probability of having at least one $^{13}$C isotope at one of the six equivalent positions is 6.4\%.  This enhanced probability and also closeness to the cancellation conditions explains the deep $^{13}$C modulation from Site G as seen in Fig.~\ref{fig:npar111-ESEEM}A.

\section{\label{sec:Conclusion} Conclusion}
In summary,  we report on electron spin coherence measurements of ensembles of NV$^-$ centers in diamond at X-band magnetic fields and low temperatures ($<70$~K).  High energy electron irradiation (to generate vacancies) and subsequent annealing (to diffuse vacancies) was used to produce NV$^-$ centers in synthetic type IIb diamonds with a nitrogen impurity concentration less than $1$~ppm. We show that the annealing step is critical in order to achieve a high yield of NV$^-$ centers and also to repair residual damage due to the electron irradiation.  Insufficient annealing (900$^\circ$C for 20 mins) leaves unrepaired damage behind resulting in faster spin decoherence for NV$^-$.  From the temperature dependence of this damage-related decoherence process, and assuming a simple thermally-activated noise model, we deduce a characteristic activation energy of 2.5 meV and density of 2.7$\cdot$10$^{16} $cm$^{-3}$ for the residual damage defects. We find that a higher temperature anneal (1000$^\circ$C for 60 mins) repairs the damage and removes the additional decoherence.  In the properly annealed diamonds T$_2$ = 0.74~ms at 5~K and is independent of field orientation, limited by spectral diffusion from a natural abundance 1.1$\%$ of  $^{13}$C nuclear spins. This is evidenced by an exponential stretch factor close to 2 in the NV$^-$ decoherence decays.
 
A strong ESEEM from distant $^{13}$C nuclei ($>0.5$~nm) is observed in low-field ODMR experiments~\cite{Glasbeek1990,VanOort1990,Maze2008,Stanwix2010}, but is fully suppressed at our X-band magnetic fields (280-400~mT).  Instead a strong ESEEM arising from the central $^{14}$N nucleus is observed.  By assuming complete $C_{3v}$ symmetry for NV$^-$, we can extract accurate $^{14}$N nuclear quadrupole and hyperfine tensors.  With the magnetic field aligned along the symmetry axis the  $^{14}$N modulation vanishes and instead we observe a modulation from proximal $^{13}$C sites (identified as 0.3 and 0.5 nm away from the NV$^-$), consistent with previous low-field ODMR experiments and density functional theory calculations.\cite{Gali2008,Smeltzer2011,Dreau2012}

\section{\label{sec:Acknowledgements} Acknowledgements}
This work was supported by the NSF and EPSRC through the Materials World Network and NSF MRSEC Programs (Grant No. DMR-1107606, EP/I035536/1, and DMR-01420541), and the ARO (Grant No. W911NF-13-1-0179).

\appendix
\section{\label{sec:Appendix} Time-Domain Traces for $^{14}$N ESEEM in NV$^-$}
\begin{figure*}[t]
\includegraphics[width=\linewidth]{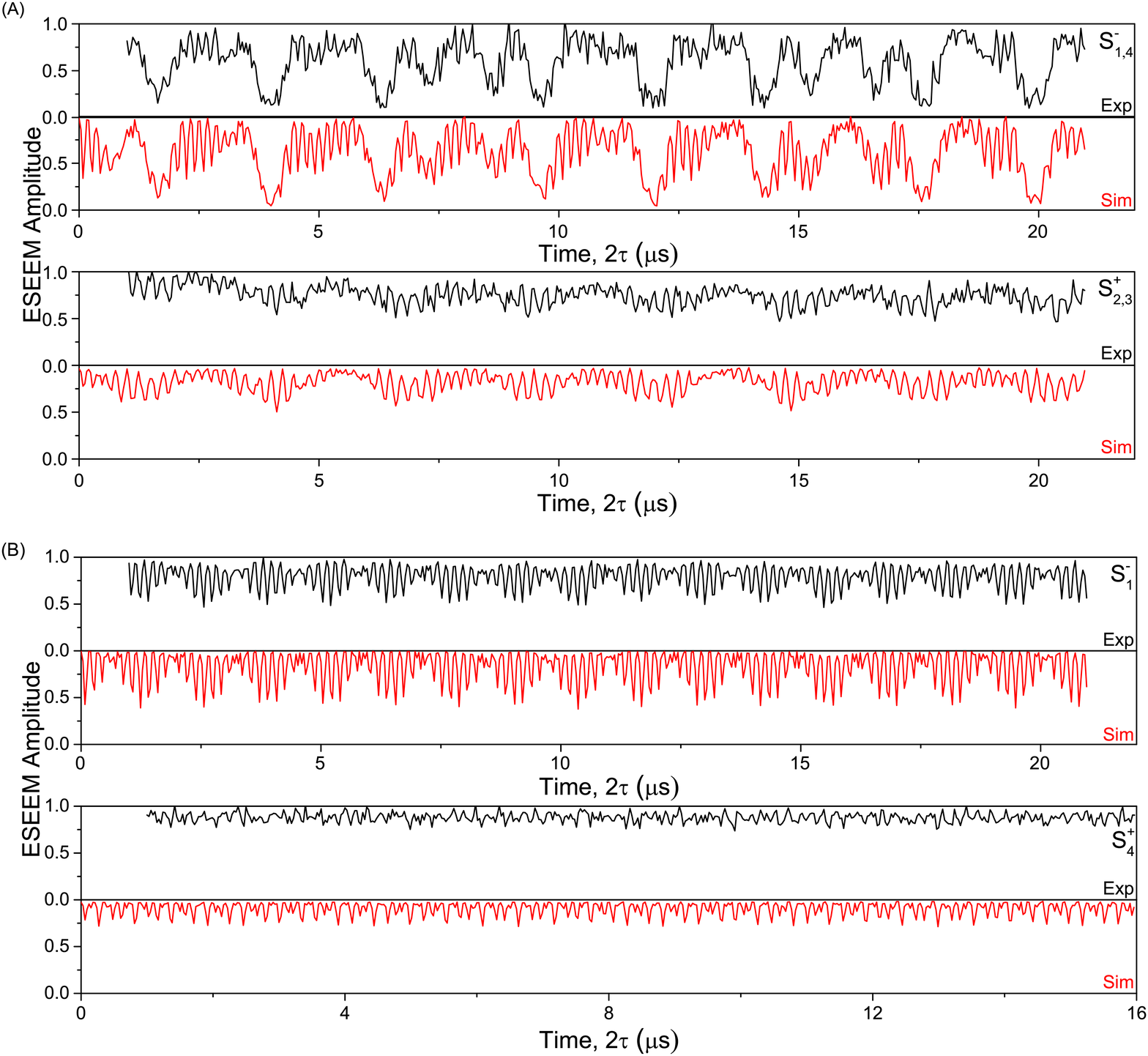}
\caption{(Color online)  Experimental (black) and simulated (red) $^{14}$N ESEEM time-domain traces in two-pulse Hahn echo experiments for NV$^-$ centers in Sample~C at 4.8~K.  Measured on (A) the ESR peaks S$^-_{1,4}$ and S$^+_{2,3}$ with B$_0\parallel [001]$, and (B) the ESR peaks S$^-_{1}$ and S$^+_{4}$ with B$_0\parallel [110]$. Experimental traces were normalized by the relaxation decay. Only the first 16-20~$\mu$s of the time-domains are shown, although the modulation extends for hundreds of microseconds with little modulation damping. Simulations were done with EasySpin~\cite{EZSpin} and using the spin Hamiltonian of Eq.~\ref{eq:ESEEMH}.}
\label{fig:TimeDomains}
\end{figure*}

The depth of $^{14}$N ESEEM modulation for NV$^-$ varies strongly as a function of magnetic field strength, orientation, and also depends on the excited ESR transition as seen in the time-domain traces in Fig.~\ref{fig:TimeDomains}. These depth variations reflect the $C_{3v}$ symmetry of the NV$^-$ center with the symmetry axis directed along [111].  The modulation depth across several orientations and transitions is accurately reproduced in our simulations, confirming the model of  Eq.~\ref{eq:ESEEMH}. 

When normalized to the overall relaxation decay (e.g. T$_2$ decay), the $^{14}$N modulation is visible for hundreds of microseconds showing no sign of modulation damping.  In ensemble experiments, a distribution of hyperfine or nuclear quadrupolar coupling parameters between the spins usually results in modulation damping. The absence of modulation damping for the case of NV$^-$ is indicative of a very narrow distribution of the parameters.  From the observed window we can estimate that $\delta \text{A/A}< 0.2\%$ and $\delta \text{Q/Q}< 1\%$ for NV$^-$ centers in our samples.

\bibliographystyle{aipnum4-1}
\bibliography{references_v13_amt}

\end{document}